\pdfoutput=1
\documentclass{IEEEtran4PSCC}
\usepackage{amsmath,amsfonts,amssymb,amsthm}

\usepackage{graphicx}
\usepackage{float}
\usepackage{caption}
\usepackage{bm} 
\usepackage{dsfont}
\usepackage{wasysym}
\usepackage{nameref}
\usepackage[symbol]{footmisc}

\usepackage{hyperref}
\usepackage{cleveref}
\newtheorem{theorem}{Theorem}[section]
\newtheorem{lemma}[theorem]{Lemma}
\newtheorem{remark}[theorem]{Remark}
\newtheorem{definition}[theorem]{Definition}
\newtheorem{corollary}[theorem]{Corollary}
\newtheorem{fact}[theorem]{Fact}

\begin{document}
	\title{Robustly Maximal Utilisation of\\Energy-Constrained Distributed Resources}
	\author{
		\IEEEauthorblockN{Michael Evans and David Angeli$^*$}
		\IEEEauthorblockA{\small{Department of Electrical and Electronic Engineering,}\\Imperial College London, U.K.\\
			\{m.evans16, d.angeli\}@imperial.ac.uk}
		\and
		\IEEEauthorblockN{Simon H. Tindemans}
		\IEEEauthorblockA{\small{Department of Electrical Sustainable Energy,}\\ TU Delft, Netherlands\\
			s.h.tindemans@tudelft.nl}
	}
	\maketitle
	\footnotetext[1]{Dr David Angeli is also with the Department of Information Engineering, University of Florence, Italy}
	
	\begin{abstract}
		We consider the problem of dispatching a fleet of distributed energy reserve devices to collectively meet a sequence of power requests over time. Under the restriction that reserves cannot be replenished, we aim to maximise the survival time of an energy-constrained islanded electrical system; and we discuss realistic scenarios in which this might be the ultimate goal of the grid operator. We present a policy that achieves this optimality, and generalise this into a set-theoretic result that implies there is no better policy available, regardless of the realised energy requirement scenario.
	\end{abstract}
	
	\section{Introduction}
	\subsection{Background}
	Recent years have seen an increasing abundance of distributed energy resources (DERs) connected to electricity grids. Many of these resources are energy-constrained, such as batteries with a limited capacity, or flexible demand that can postpone power consumption within limits set by users. Collectively, such resources can provide a range of valuable services to the system, such as supporting the real-time balancing of supply and demand. System operators are offering frameworks for the commercial delivery of such services, e.g. the California Independent System Operator (CAISO) \cite{CAISO} and the Pennsylvania-New Jersey-Maryland Interconnection RTO (PJM) \cite{PJM} in the US. We consider a control framework as exemplified by PJM, which broadcasts a system-wide regulation signal that is to be tracked. Note that by considering such a regulation signal, we are able to generalise our approach to cover the full range of network configurations, including the pre-existence of distributed intermittent generation or additional demand response. This approach represents a centralised control of the \emph{aggregate} response of a collection of resources, which for the purposes of description we assume to be delivered by an aggregator. We assume that there is an agreement between the network operator and the aggregator whereby the amount of power requested is updated at regular intervals and the aggregator must make a decision as to which of its resources to deploy to meet that request. We consider the aggregator to be offering the service of system support during periods of supply shortfall, and do not restrict our devices to perfect efficiency. We also assume the absence of cross charging between devices, which corresponds to a regime in which operational losses are minimised. Hence we restrict all devices to discharging operation only.
	
	Within this contractual framework we focus on the decision making of the aggregator. There has been significant literature to date on aggregator-based control approaches, across a wide range of devices. These vary from electric vehicles \cite{Wenzel2017} to thermostatically controlled loads (TCLs) \cite{Tindemans2015,Vrettos2016} and pool pumps \cite{Meyn2014}. Of particular relevance to this paper are considerations of how best to dispatch devices in the presence of uncertain demand. This has been explicitly considered in \cite{Bejan2012,Gast2014,Cruise2015}, applied in the latter to arbitrage as well as buffering. Within the field of DER allocation more specifically, prior work has taken into account uncertainty in the form of achieving optimality in expectation \cite{Alharbi2015,Saber2012}. Other authors have considered robust approaches that apply a minimax optimality condition in order to achieve the smallest cost of operation across all possible scenarios \cite{Bai2015}. In addition, authors have investigated ways of obtaining probabilistic capacity-duration relationships, as a means of capturing the uncertainty present \cite{Zhang2017}, and demonstrated the effects of such uncertainty on the actual capabilities of distributed resources \cite{Mathieu2013}. We take a more general approach, aiming to maximise the set of power requests in the future, currently unknown to the aggregator, that can be successfully satisfied by the fleet. Under the restriction that devices are able to discharge only, we consider the problem of maximising time to failure of the fleet in the absence of any information about future request signals. To this end, the contribution of this work is twofold. Our approach firstly yields the result that a maximal solution of a set-inclusion form exists. Secondly, we present an explicit feedback policy and show that this achieves optimality in this set-theoretic sense. Hence, we show that optimality holds regardless of the actual request profile that the aggregator receives.		
	
	The remainder of this paper is organised as follows. Section~\ref{sec: prob formulation} describes the motivation behind this research and the modelling of the problem. Section~\ref{sec: op feedback} presents our explicit policy and results relating to its optimality (note that proofs of the Lemmas and Theorem can be found in the \nameref{sec:appendix}). This is followed by an interpretation of these results in Section~\ref{sec: interp}. Comparative policies are introduced in Section~\ref{sec:sim}, followed by representative simulation results. Finally, Section~\ref{sec:conc} concludes the report and discusses relevant future work.
	
	\section{Problem Formulation}
	\label{sec: prob formulation}
	
	\subsection{Assumptions}	
	The problem framework within which our investigation takes place can be summarised as follows:
	\begin{enumerate}
		\item	A resource controller has an ensemble of devices available to it at the starting time.
		\item	These devices are energy constrained.
		\item	The resource controller has agreed to comply with an externally imposed reference signal, if technically feasible.
		\item	They have no prior information about the requests they will receive, and must instantaneously meet any request (communication and calculation delays are assumed negligible for demonstration).
		\item	If the demand persists prior to replenishing the available energy, it is inevitable that at some point in the future they will be unable to meet a request, given their finite reserves. However, they would like to postpone this output shortfall as far as possible into the future. We take this to be the cost functional in our optimisation problem: maximise time to failure.
	\end{enumerate}

	\subsection{Motivation}
	
	The assumptions above are reflective of a number of real-world scenarios, albeit in stylised form. A straightforward example is that of a physically isolated network with a finite supply of energy, e.g. diesel generators on an offshore platform. If the actual consumption pattern is unpredictable to some extent, it is prudent to maximise the `mission time' of the network by dispatching scarce resources in an optimal way. 
	
	In a more general sense, the assumptions can reflect the problem faced by a network operator after the occurrence of an event that threatens security of supply. The operator's objective is to meet demand on the network for as long as possible until the event naturally finishes or more permanent mitigation can be implemented. For example, when an unexpectedly large peak in demand or shortfall in wind output is experienced, purely surviving until the end of this event would be sufficient to maintain normal grid operation; and, in this case, the grid operator should employ any battery resources it has available to it according to the optimal policy that we propose. As a second example, when an islanded microgrid is formed due to fault conditions, either in the case of a microgrid that is tied to the main network losing this connection or the unexpected islanding of a portion of the main grid, the objective of the grid operator would be to maintain functionality of the islanded microgrid up to the point in time at which reconnection could be achieved. Both scenarios result in a situation where the grid operator would like to maintain operation for as long as possible given its newly-limited reserves, in line with the assumptions outlined above. 
	
	In the following, we present a policy for energy-limited distributed resources that maximises the survival time (time before forced outages or incurring non-compliance penalties). From this point on, in the interests of clarity, we will discuss the dispatch decisions as being undertaken by an aggregator. Note, however, that in scenarios where significant loss of load is at stake, the grid operator itself may take on the role of central dispatcher.
	
	\subsection{Mathematical formulation}
	\noindent Given this set-up, we denote by $n$ the number of DER devices available to the aggregator, and define the set of all devices as follows:
	\begin{equation}
	\mathcal{N}\doteq\{D_1,D_2,...,D_n\}.
	\end{equation}
	
	\noindent We model these devices as energy-constrained generating units, for example generators with limited fuel stored or batteries that discharge only (during the time frame of interest). We do not impose any restrictions on homogeneity of devices and allow each device to have a unique discharging efficiency $\eta_i$. For convenience we incorporate this into the model implicitly by considering the extractable energy of a device,
	\begin{equation}
	E_i(t)\doteq \eta_iE^s_i(t),
	\end{equation}
	in which $E^s_i(t)$ denotes the total stored energy by device $D_i$ at time $t$. From this point onwards we will solely consider the extractable energy, referred to simply as the energy. We choose the power delivered by each device to be the control input $u_i(t)$, and assume that this is measured externally so that efficiency is once again accounted for. This leads to integrator dynamics on the energy of each device
	\begin{equation}
	\dot{E}_i(t)=-u_i(t),
	\end{equation}
	subject to the physical constraint
	\begin{equation}
	E_i(t)\geq0.
	\end{equation}
	
	\noindent Resulting from the choice of exclusive discharging ability, the power of each device is constrained as
	\begin{equation}
	u_i(t)\in[0,\bar{p}_i],
	\end{equation}	
	in which $\bar{p}_i$ denotes the maximum discharge rate of device $D_i$, and with the  convention that discharging rates are positive. Note, once again, that there is no homogeneity imposed on energy or power constraints across the ensemble of devices. We define the \textit{time-to-go} of device $D_i$ to be the time remaining for which this device can run at its maximum power, i.e.
	\begin{equation}
	x_i(t) \doteq \frac{E_i(t)}{\bar{p}_i},
	\end{equation}	
	and represent the state of each device by its time-to-go. We then form a state vector by stacking this value across all the devices
	\begin{equation}
	x(t)\doteq \begin{bmatrix}
	x_1(t)\ \dots\ x_n(t)
	\end{bmatrix}^T,
	\end{equation}	
	and define the state space
	\begin{equation}
	\mathcal{X}\doteq [0,+\infty)^n.
	\end{equation}
	
	\noindent We similarly stack the control inputs across devices to give a vector input
	\begin{equation}
	u(t) \doteq \begin{bmatrix}
	u_1(t)\ \dots\ u_n(t)
	\end{bmatrix}^T.
	\end{equation}
	
	\noindent For clarity we employ the notation
	\begin{equation}
	\bar{p}\doteq\begin{bmatrix}
	\bar{p}_1\ \dots\	\bar{p}_n
	\end{bmatrix}^T,
	\end{equation}	
	and form the product set of our constraints on all the inputs,
	\begin{equation}
	\mathcal{U}_{\bar{p}} \doteq [0,\bar{p}_1] \times [0,\bar{p}_2] \times ... \times [0,\bar{p}_n],
	\end{equation}	
	so that we can compactly write our input constraints as
	\begin{equation}
	u(t)\in\mathcal{U}_{\bar{p}}.
	\end{equation}
	
	\noindent We also form a diagonal matrix of maximum powers,
	\begin{equation}
	P \doteq \text{diag}(\bar{p}),
	\end{equation}
	so that our dynamics can be written in matrix form as
	\begin{equation}
	\dot{x}(t)=-P^{-1}u(t).
	\end{equation}
	
	\noindent  We denote by $P^r(\cdot)$ a power reference signal received by the aggregator, and in addition denote a truncated trajectory of such a signal as
	\begin{equation}
	P^r_{[0,t)}\doteq\left\{
	\begin{array}{@{}ll@{}}
	P^r(\tau), & \text{if}\ \tau\in[0,t) \\
	0, & \text{otherwise}.
	\end{array}\right.
	\end{equation} 
	For any reference signal to be \textit{feasible}, it needs to be admissible for all time; in other words satisfiable without violating any constraints. We define the set of such signals as follows:
	\begin{definition}
		The set of \textbf{feasible} power reference signals for a system of maximum powers $\bar{p}$ in state $x$ is defined as
		\begin{equation*}
		\begin{aligned}
		\mathcal{F}_{\bar{p},x} \doteq \bigg\{P^r : [0,+\infty) \to &[0,+\infty)\ \big|\ \exists u(\cdot)\ \big|\ \ \forall t\geq 0, \\
		u(t)&\in \mathcal{U}_{\bar{p}}, \\
		1^Tu(t)&=P^r(t), \\
		\dot{x}(t)&=-P^{-1}u(t), \\
		x(0)&=x, \\
		x_i(t) &\geq 0,\ \ i=1,2,...,n \hspace{1cm}\bigg\},
		\end{aligned}
		\end{equation*}
		in which $1$ denotes the unity vector of appropriate length, $n$.
	\end{definition}
	
	\noindent We say that the \emph{future flexibility} of a system is larger than that of another system when its feasible set is strictly greater. Any input trajectory satisfying a feasible reference will result in a state trajectory that \textit{fulfils} said reference, defined as follows:
	\begin{definition}
		A state trajectory $z(\cdot)$ can be said to \textbf{fulfil} a power reference signal $P^r(\cdot)$ if
		\begin{equation*}
		\begin{aligned}
		\exists u(\cdot)&\colon\ \forall t\geq 0, \\
		u(t)&\in \mathcal{U}_{\bar{p}}, \\
		1^Tu(t)&=P^r(t), \\
		\dot{z}(t)&=-P^{-1}u(t), \\
		z(0)&=x, \\
		z_i(t) &\geq 0,\ \ i=1,2,...,n.
		\end{aligned}
		\end{equation*}		
	\end{definition}

	\noindent For comparison between different state trajectories, it is also useful to denote the maximum instantaneous power available in state $x$ as $\bar{P}^r(x)$, which we define as follows:
	\begin{equation}
	\bar{P}^r(x)\doteq\sum_{i\colon x_i>0}\bar{p}_i\ .
	\end{equation}
	
	\noindent We consider the scenario in which the ability to satisfy the power request takes precedence over other objectives. Note that in an economic context this is equivalent to associating a very high cost with a failure to meet demand. This assumption allows us to consider the concept of survival time maximisation in a general sense, without considering price dynamics in detail. We denote \emph{time to failure} as $\Theta_{\bar{p},x}(P^r)$ and define it as follows:
	\begin{equation}
	\Theta_{\bar{p},x}(P^r) \doteq \text{sup}\big\{t\ \vert\ P^r_{[0,t)}\in\mathcal{F}_{\bar{p},x}\big\} .
	\end{equation}
	
	\noindent Our interpretation of time to failure as being the objective function in our optimisation allows us to compare two device configurations. We are then able to generalise this comparison into a more powerful set-theoretic maximisation as follows:
	\begin{fact}
		\label{fact:set-theoretic max}
		Given the system-state pair $(\bar{p}^a,x^a)$ with at least as great a future flexibility as the system-state pair $(\bar{p}^b,x^b)$,
		\begin{equation*}
		\mathcal{F}_{\bar{p}^a,x^a}\supseteq	\mathcal{F}_{\bar{p}^b,x^b} \implies \Theta_{\bar{p}^a,x^a}(P^r) \geq \Theta_{\bar{p}^b,x^b}(P^r)\ \ \ \forall P^r,
		\end{equation*}
		and hence the time to failure of $(\bar{p}^a,x^a)$ is at least as great as that of $(\bar{p}^b,x^b)$, under any reference.
	\end{fact}
	
	\section{Optimal Feedback Policy}
	\label{sec: op feedback}
	We present the following explicit feedback law. Without loss of generality, reorder the states by descending value and group them into collections of equal value (leading to $q$ such groups),
	\begin{subequations}
		\label{eq: explicit policy}
		\begin{equation}
		\begin{aligned}
		x_1=x_2=...=x_{s_1} &> x_{s_1+1}=...=x_{s_2} \\
		&>... > x_{s_{q-1}+1}=...=x_{s_q} .
		\end{aligned}
		\end{equation}
		
		\noindent Additionally, form stacked vectors of the inputs and maximum powers respectively across the devices in each group,
		\begin{equation}
		U_i\doteq \begin{bmatrix}
		u_{s_{i-1}+1} \\
		\vdots \\
		u_{s_i}
		\end{bmatrix},\ \ \ \bar{U}_i \doteq \begin{bmatrix}
		\bar{p}_{s_{i-1}+1} \\
		\vdots \\
		\bar{p}_{s_i}
		\end{bmatrix},
		\end{equation}	
		with the convention that $s_0=0$. The explicit feedback law is then calculated as a fraction $r_i$ of the maximum power $\bar{U}_i$ according to:
		\renewcommand{\arraystretch}{1.5}
		\begin{equation}
		r_i=\left\{
		\begin{array}{@{}ll@{}}
		1, & \text{if}\ \sum_{j\leq i}1^T\bar{U}_j\leq P^r \\
		0, & \text{if}\ \sum_{j< i}1^T\bar{U}_j\geq P^r \\
		\frac{P^r-\sum_{j< i}1^T\bar{U}_j}{1^T\bar{U}_i}, & \text{otherwise},
		\end{array}\right. \\
		\end{equation}
		\begin{equation}
		\kappa(x,P^r)\doteq\begin{bmatrix}
		U_1^T\ \dots\ U_q^T
		\end{bmatrix}^T\ \ \textnormal{with}\ \  U_i=r_i\bar{U}_i .
		\end{equation}
	\end{subequations}
	\renewcommand{\arraystretch}{1}
	\hspace{-2.2mm} Denoting by $z^*(\cdot)$ the state trajectory under the application of \eqref{eq: explicit policy}, the closed-loop dynamics are then
	\begin{equation}
	\label{eq:closed-loop dynamics}
	\dot{z}^*(t)=-P^{-1}\kappa\big(z^*(t),P^r(t)\big) .
	\end{equation}
	
	\noindent We will now present relevant results and remarks relating to this feedback law. Note that for clarity of argument the proofs of Lemmas~\ref{Lemma III.1} and \ref{Lemma III.5} and Theorem~\ref{Theorem III.7} can be found in the \nameref{sec:appendix}.
	
	\begin{lemma}\label{Lemma III.1}
		A unique solution of \eqref{eq:closed-loop dynamics} exists for all initial conditions.
	\end{lemma}
	\newcommand{\proofExistence}{		
		\noindent Let $\mathcal{P}$ be the set of all partitions of $\mathcal{N}$, and for convenience of notation denote a single partition as
		\begin{subequations}	
			\label{eq:partitionSubsets}		
			\begin{equation}
			p=\big\{\mathcal{N}_1,\mathcal{N}_2,...,\mathcal{N}_k\big\},
			\end{equation}
			\begin{equation}
			p\in\mathcal{P}.
			\end{equation}		
		\end{subequations}
		
		\noindent A corresponding subset of state-space $\mathcal{X}_p\in\mathcal{X}$ can be introduced as
		\begin{equation}
		\begin{aligned}
		\mathcal{X}_p\doteq\bigg\{&x\in\mathcal{X}\ \big|\ \forall i,j\in\{1,...,n\},\ x_i=x_j \\
		&\iff \exists l\ \big|\ D_i\in \mathcal{N}_l\ \&\ D_j\in\mathcal{N}_l\bigg\}.
		\end{aligned}
		\end{equation}
		
		\noindent Note that all such state-subspaces form a partition of $\mathcal{X}$. In particular
		\begin{equation}
		\mathcal{X}=\bigcup_{q\in\mathcal{P}}\mathcal{X}_q\ .
		\end{equation}
		
		\noindent Denote by $t_0$ the time at which the state enters the subspace $\mathcal{X}_p$. Similarly denote by $\bar{t}_{x(t_0)}$ the time at which it leaves this subspace,
		\begin{equation}
		\bar{t}_{x(t_0)}\doteq\text{sup}\ \big\{s\geq t_0\ \big|\ x(t)\in\mathcal{X}_p\ \forall t\in[t_0,s]\big\}.
		\end{equation}
		
		\noindent Now, for any $x(t_0)\in\mathcal{X}_p$ and any $t\in[t_0,\bar{t}_{x(t_0)}]$, define
		\begin{equation}
		\begin{aligned}
		x_i(t)=&x_i(t_0) \\
		&-\int_{t_0}^t\text{max}\bigg\{\text{min}\bigg\{\frac{P^r(\theta)-\sum_{D_j\in\mathcal{N}_{k<l}}\bar{p}_j}{\sum_{D_m\in\mathcal{N}_l}\bar{p}_m},1\bigg\},0\bigg\}d\theta \\
		&\hspace{5.5cm}\forall D_i\in\mathcal{N}_l,\  \forall l.
		\end{aligned}
		\label{eq: subspace evolution}
		\end{equation}
		
		\noindent Consider the partitioning of devices as the state evolves across $\mathcal{X}_p$ according to \eqref{eq: subspace evolution}. For each subset $\mathcal{N}_l$, \eqref{eq: subspace evolution} ensures that all devices in the subset are instantaneously left idle, run at full power or at an equal fraction thereof. In each case, the device state values move in unison and so $x(t)$ belongs to $\mathcal{X}_p$ until two or more groups of devices reach the same time-to-go. We informally refer to such an event as an \textit{equalisation}. There must be a finite positive time between equalisations, hence
		\begin{equation}
		\bar{t}_{x(t_0)} > t_0.
		\end{equation}
		
		\noindent Considering right derivatives, i.e.
		\begin{equation}
		\dot{x}(t)\doteq \underset{\ h\to0^+}{\lim}\frac{x(t+h)-x(t)}{h},
		\end{equation}
		and having shown that the partition is maintained, we are able to compute the dynamics across one subspace $\mathcal{X}_p$ as
		\begin{equation}
		\begin{aligned}
		\dot{x}_i(t)=&-\text{max}\bigg\{\text{min}\bigg\{\frac{P^r(t)-\sum_{D_j\in\mathcal{N}_{k<l}}\bar{p}_j}{\sum_{D_m\in\mathcal{N}_l}\bar{p}_m},1\bigg\},0\bigg\}\\
		&\hspace{4.5cm}\forall D_i\in\mathcal{N}_l,\  \forall l.
		\end{aligned}
		\label{eq: maximin dynamics}
		\end{equation}
		
		\noindent Thus \eqref{eq: subspace evolution} is a solution to \eqref{eq: maximin dynamics} over the open interval $t\in [t_0,\bar{t}_{x(t_0)})$. The dynamics of (\ref{eq: maximin dynamics}) correspond to the implementation of the policy (\ref{eq: explicit policy}), therefore this policy results in a continuous state trajectory as in (\ref{eq: subspace evolution}) across the same time interval. We are then able to recognise a unique continuous solution by composing these trajectories across all state-subspaces. 
		
		Note that when solutions enter a new element $\mathcal{X}_p$ of the partition of $\mathcal{X}$, the number of strict inequalities characterising such a region strictly decreases. Hence at most $n$ switches occur along any given trajectory, so there is no possibility of an accumulation point of switches between subspaces. Therefore a unique solution of the closed-loop dynamics \eqref{eq:closed-loop dynamics} must exist.
	}
	
	\begin{remark}
		Under the application of \eqref{eq: explicit policy}, an increasing number of devices is utilised up to whichever occurs earlier: all available devices are utilised or the reference is fulfilled. It is therefore trivial to see that
		\begin{equation}
		1^T\kappa\big(x,P^r\big)=P^r \iff P^r\leq \bar{P}^r
		\end{equation}
		according to this policy. 
	\end{remark}
	
	\begin{remark}
		The subset orders are preserved under the feedback policy \eqref{eq: explicit policy}, i.e., without loss of generality,
		\begin{equation}
		\begin{aligned}
		&x_{s_1}(0) \geq ... \geq x_{s_q}(0) \\
		&\implies x_{s_1}(t) \geq ... \geq x_{s_q}(t)\ \ \forall t\geq0.
		\end{aligned}
		\end{equation}
	\end{remark}
	\begin{remark}
		Under the feedback policy \eqref{eq: explicit policy}, devices with the same initial time-to-go will be run at equal fractions of their maximum powers forever.
	\end{remark}
	\noindent Utilisation of these remarks leads to the following result:
	\begin{lemma}
		\label{Lemma III.5}
		Let $P^r(\cdot)$ be a feasible reference and $\tilde{z}(\cdot)$ a solution that fulfils $P^r(t)$, for all $t\geq0$. Then the solution $z^*(\cdot)$ arising from feedback \eqref{eq: explicit policy} and initialised as $z^*(0)=\tilde{z}(0)$ 
		is well-defined for all $t\geq0$, and moreover:
		\begin{equation}
		\textnormal{supp}[z^*(t)]\supseteq\textnormal{supp}[\tilde{z}(t)]\ \ \ \forall t\geq0 ,
		\end{equation}
		in which $\textnormal{supp}[x(t)]$ denotes the support of state $x(t)$, interpretable as the set of devices that are not empty at time $t$.
	\end{lemma}
	\newcommand{\proofSupport}{
		
		We first compose a framework as follows. Consider the partition $p$, induced at the initial time by $\tilde{z}(0)~\in~\mathcal{X}_p$, and let
		\begin{equation}
		\mathcal{N}_l,\ l=1,2,...,q
		\end{equation}
		be the corresponding subsets as in \eqref{eq:partitionSubsets}. Then consider, for an arbitrary positive integer $r$, the family $\mathcal{Q}$ obtained by joining the $\mathcal{N}_l$s corresponding to the devices with the $r$ lowest (distinct) values of time-to-go:
		\begin{equation}
		\mathcal{Q}_r\doteq\mathcal{N}_{q-r+1}\cup...\cup\mathcal{N}_q .
		\end{equation}
		In general, we know that the policy \eqref{eq: explicit policy} chooses to run the devices contained in $\mathcal{Q}_r$ as little as possible, and only then in the case when devices with higher times-to-go are run at maximum power. Utilising the notation that $z_{\mathcal{Q}_r}$ denotes the $z$-vector truncated to the devices which are elements of $\mathcal{Q}_r$, and likewise that $1_{\mathcal{Q}_r}$ and $P_{\mathcal{Q}_r}$ denote the unity vector and $P\text{-matrix}$ truncated to the corresponding elements, we are able therefore to say that
		\begin{equation}
		1_{\mathcal{Q}_r}^TP_{\mathcal{Q}_r}\dot{z}^*_{\mathcal{Q}_r}(t) \geq 	1_{\mathcal{Q}_r}^TP_{\mathcal{Q}_r}\dot{\tilde{z}}_{\mathcal{Q}_r}(t)\ \ \forall t\geq0, \forall r.
		\label{eq: Q set}
		\end{equation}
		Now, denote by $\mathcal{Q}_{r_0}$ the set of devices empty at time $t=0$. In addition, let $\bar{t}_1$ denote the first time at which one or more additional devices are emptied along the $z^*(\cdot)$ solution:
		\begin{equation}
		\bar{t}_1 \doteq \textnormal{inf}\ \big\{t\geq0\ \vert\ \exists D_i\in\mathcal{N}\ \vert\ z^*_i(t)=0,\ z^*_i(0)>0\big\}.
		\end{equation}
		
		\noindent Consider the open interval $[0,\bar{t}_1)$. By definition, no devices are emptied along the $z^*(\cdot)$ solution over this interval, hence
		\begin{equation}
		\begin{aligned}
		\text{supp}[z^*(\tau)]&=\mathcal{N}\setminus\mathcal{Q}_{r_0}=\text{supp}[\tilde{z}(0)]\\
		&\supseteq\text{supp}[\tilde{z}(\tau)]\ \ \ \forall \tau\in[0,\bar{t}_1),
		\end{aligned}
		\end{equation}
		where the inclusion follows because the support along the $\tilde{z}(\cdot)$ solution is non-increasing. In addition, since $\tilde{z}$ fulfils $P^r$, then
		\begin{equation}
		P^r(\tau)\leq\sum_{D_i\notin\mathcal{Q}_{r_0}}\bar{p}_i\ \ \forall \tau\in[0,\bar{t}_1),
		\end{equation}
		and so $z^*$ is well-defined and non-negative over the open interval $[0,\bar{t}_1)$. 
		
		Now consider the time instant $t=\bar{t}_1$. By continuity of $z^*$,
		\begin{equation}
		z_i^*(\bar{t}_1)=0,
		\end{equation}
		in which the index $i$ refers to the one or more devices that are emptied at exactly this time. Moreover, because
		\begin{equation}
		z_i^*(0)=z_j^*(0)\implies z_i^*(t)=z_j^*(t)\ \ \ \forall t\geq0,
		\end{equation}
		we are able to say that
		\begin{equation}
		\begin{aligned}
		\exists r_1\colon\ \ &z_i^*(\bar{t}_1)=0\ \ \ \forall D_i\in\mathcal{Q}_{r_1}, \\
		&z_k^*(\bar{t}_1)>0\ \ \ \forall D_k\notin\mathcal{Q}_{r_1}.
		\end{aligned}
		\end{equation}
		Now, consider the evolution of the trajectories $z^*_{\mathcal{Q}_{r_1}}$ and $\tilde{z}_{\mathcal{Q}_{r_1}}$. From a common initialisation,
		\begin{equation}
		1_{\mathcal{Q}_{r_1}}^TP_{\mathcal{Q}_{r_1}}z^*_{\mathcal{Q}_{r_1}}(0) = 	1_{\mathcal{Q}_{r_1}}^TP_{\mathcal{Q}_{r_1}}\tilde{z}_{\mathcal{Q}_{r_1}}(0) ,
		\end{equation}
		and so \eqref{eq: Q set} gives that
		\begin{equation}
		\begin{aligned}
		1_{\mathcal{Q}_{r_1}}^TP_{\mathcal{Q}_{r_1}}&z^*_{\mathcal{Q}_{r_1}}(t) =1_{\mathcal{Q}_{r_1}}^TP_{\mathcal{Q}_{r_1}}z^*_{\mathcal{Q}_{r_1}}(0)\\
		&\hphantom{z^*_{\mathcal{Q}_{r_1}}(t) =}+\int_0^{t}1_{\mathcal{Q}_{r_1}}^TP_{\mathcal{Q}_{r_1}}\dot{z}^*_{\mathcal{Q}_{r_1}}(\tau)d\tau \\
		&\geq1_{\mathcal{Q}_{r_1}}^TP_{\mathcal{Q}_{r_1}}\tilde{z}_{\mathcal{Q}_{r_1}}(0)+\int_0^t1_{\mathcal{Q}_{r_1}}^TP_{\mathcal{Q}_{r_1}}\dot{\tilde{z}}_{\mathcal{Q}_{r_1}}(\tau)d\tau \\
		&=1_{\mathcal{Q}_{r_1}}^TP_{\mathcal{Q}_{r_1}}\tilde{z}_{\mathcal{Q}_{r_1}}(t)\ \ \ \forall t\geq0.
		\end{aligned}
		\end{equation}
		Hence
		\begin{equation}
		z^*_{\mathcal{Q}_{r_1}}(\bar{t}_1)=0\implies \tilde{z}_{\mathcal{Q}_{r_1}}(\bar{t}_1)=0,
		\end{equation}
		and moreover
		\begin{equation}
		\textnormal{supp}[z^*(\bar{t}_1)]\supseteq\textnormal{supp}[\tilde{z}(\bar{t}_1)].
		\end{equation}
		In addition, since 
		$\tilde{z}$ fulfils $P^r$, then
		\begin{equation}
		P^r(\bar{t}_1)\leq\sum_{D_i\notin\mathcal{Q}_{r_1}}\bar{p}_i,
		\end{equation}		
		and so $z^*$ is well-defined and non-negative over the closed interval $[0,\bar{t}_1]$. Now, to extend this logic further into the future, denote by $\bar{t}_i$ the time at which the $i^\textnormal{th}$ emptying event occurs, defined for $i>1$ as
		\begin{equation}
		\bar{t}_i \doteq \textnormal{inf}\ \big\{t>\bar{t}_{i-1}\ \vert\ \exists D_j\in\mathcal{N}\ \vert\ z^*_j(t)=0,\ z^*_j(\bar{t}_{i-1})>0\big\}.
		\end{equation}
		Equivalent logic to above leads to at least the result that
		\begin{equation}
		\begin{aligned}
		&\textnormal{supp}[z^*(\bar{t}_i)]\supseteq\textnormal{supp}[\tilde{z}(\bar{t}_i)]\\
		&\hspace{0.8cm}\implies\textnormal{supp}[z^*(\bar{t}_{i+1})]\supseteq\textnormal{supp}[\tilde{z}(\bar{t}_{i+1})].
		\end{aligned}
		\end{equation}
		Hence, by induction,
		\begin{equation}
		\textnormal{supp}[z^*(t)]\supseteq\textnormal{supp}[\tilde{z}(t)]\ \ \ \forall t\geq0.
		\end{equation}	
	}
	\noindent This result then leads to the following Corollary:
	\begin{corollary}
		\label{cor: max inst power}
		The policy \eqref{eq: explicit policy} maximises the instantaneous power request that can be met by the system, i.e.
		\begin{equation}
		\bar{P}^r\big(z^*(t)\big)\geq\bar{P}^r\big(\tilde{z}(t)\big)\ \ \ \forall \tilde{z}(\cdot),\forall t\geq0.
		\end{equation}	
	\end{corollary}
	\noindent Arguments making use of these preceding results finally lead to the main Theorem that we present in this paper:
	
	\begin{theorem}
		\label{Theorem III.7}
		The policy \eqref{eq: explicit policy} maximises future flexibility.
	\end{theorem}
	\newcommand{\proofMaxFlex}{
		Consider a feasible reference $P^r(\cdot)$. Denote by $z^*(t)$ the state resulting from the application of (\ref{eq: explicit policy}) in the interval $[0,t]$ that meets this reference, and likewise denote by $\tilde{z}(t)$ any other feasible solution. Starting from this state, $\tilde{z}(t)$ at time $t$, pick any feasible reference, $\hat{P}^r(\cdot)\in\mathcal{F}_{\bar{p},\tilde{z}(t)}$. This results in some trajectory $\hat{z}(\cdot)$ which we define as below
		\begin{equation}
		\hat{z}(\tau)=\left\{
		\begin{array}{@{}ll@{}}
		\tilde{z}(\tau), & \text{if}\ \tau\in[0,t] \\
		\check{z}(\tau), & \text{if}\ \tau>t. \\
		\end{array}\right.
		\end{equation}
		Denote by $\hat{z}^*(\cdot)$ the trajectory obtained via the application of (\ref{eq: explicit policy}) corresponding to the reference
		\begin{equation}
		P^r \APLlog \hat{P}^r(\cdot) ,
		\end{equation}
		in which the $\APLlog$ operator denotes the concatenation of two signals. Resulting from Lemma \ref{Lemma III.5}, we know that
		\begin{equation}
		\textnormal{supp}[\hat{z}^*(\tau)]\supseteq\textnormal{supp}[\hat{z}(\tau)]\ \ \ \forall \tau\geq0 .
		\end{equation}
		Hence,
		\begin{equation}
		\bar{P}^r\big(\hat{z}^*(\tau)\big)\geq\bar{P}^r\big(\hat{z}(\tau)\big)\geq P^r(\tau)\ \ \ \forall \tau\geq0,
		\end{equation}
		and therefore the power constraints are satisfied along the solution $\hat{z}^*$, and so 
		\begin{equation}
		\hat{P}^r\in\mathcal{F}_{\bar{p},z^*(t)}.
		\end{equation}
		Therefore
		\begin{equation}
		\mathcal{F}_{\bar{p},z^*(t)}\supseteq\mathcal{F}_{\bar{p},\tilde{z}(t)},
		\end{equation}
		hence the policy (\ref{eq: explicit policy}) maximises future flexibility.		
	}
	
	\noindent Fact~\ref{fact:set-theoretic max} then allows us to deduce the following Corollaries:
	\begin{corollary}
		The policy \eqref{eq: explicit policy} maximises time to failure, under any reference $P^r$.
	\end{corollary}
	\begin{corollary}
		As the proposed policy is greedy and optimal for any possible reference, it is the best choice, regardless of the future reference signal.
	\end{corollary}
	
	\section{Interpretation of the Results}
	\label{sec: interp}
	In this section we attempt to gain an improved insight into the relevance of the results from the point of view of a system operator or aggregator. Firstly, it is worth mentioning the reasons behind the advantages that the proposed policy offers. As a result of Lemma~\ref{Lemma III.5} and Corollary~\ref{cor: max inst power}, we know that this policy maintains maximal availability of devices, and moreover that this is equivalent to the ability to satisfy the maximal instantaneous power. In fact, so long as the total time for which the instantaneous power request is above $\sum_{i=1}^{n-1}\bar{p}_i$ is no greater than the smallest initial time to go (i.e. $x_n(0)$) then all devices will remain available up to the time-to-failure of the optimal policy, identically the time instant at which all devices are depleted under this policy. Under other policies, for example those discussed in the following section, depletion of devices may occur earlier than this optimal time, as a result of which there might be insufficient devices available to meet an earlier reference value. The major advantage of utilising the proposed policy is this ability to meet the request for as long as possible, thereby providing the best prospect of being able to outlast a shortfall event such as a lack of wind output or an islanding failure.
	
	\section{Numerical Demonstrations}
	\label{sec:sim}
	\subsection{Alternative policy choices}
	Having presented analytical results which hold in general, we now additionally present numerical results to illustrate optimal discharging behaviour. To this end, in addition to the previously described optimal policy (OP) we consider alternative policy choices that one might implement; for a representative comparison we only consider greedy policies. These are detailed as follows.
	
	\subsubsection{Lowest Power First}
	Given that it is power spikes in the future that we aim to successfully fulfil, one might intuitively consider running those devices with the lowest maximum powers first, thereby attempting to maintain devices with higher maximum powers in reserve. We denote this policy Lowest Power First (LPF). It can be enacted by ordering the devices by maximum power, without loss of generality leading to
	\begin{equation}
	\bar{p}_1\leq\bar{p}_2\leq...\leq\bar{p}_n,
	\label{eq:LPF}
	\end{equation}
	and allocating devices in order from $D_1$ to $D_n$ as
	\begin{equation}
	u_i=\mathds{1}\{x_i>0\}\cdot\text{min}\bigg\{\bar{p}_i,P^r-\sum_{j<i}u_j \bigg\},
	\end{equation}	
	in which $\mathds{1}\{\cdot\}$ denotes the indicator function. Note that the choice of allocation between devices of equal maximum powers is made arbitrarily.
	
	\subsubsection{Proportion of Power}
	For an improved allocation based on comparing the maximum powers across devices, we consider the policy whereby devices are allocated by their proportion of $\bar{P}^r(x)$, the net instantaneous power available. We denote this policy Proportion of Power (PoP). In this case, no ordering is required and each device is run according to
	\begin{equation}
	u_i=\mathds{1}\{x_i>0\}\cdot\text{min}\bigg\{\bar{p}_i,\frac{\bar{p}_iP^r}{\bar{P}^r}\bigg\}.
	\end{equation}
	
	\subsection{Results}
	We now demonstrate the optimality of the proposed policy through comparison to the aforementioned alternative choices. We generate a scenario in which there are 1000 devices with both initial times-to-go and maximum powers uniformly distributed, as $x_i~\sim~U(0,10)~h$ and $\bar{p}_i~\sim~U(0,1.5)~kW$ respectively. In an attempt to model a realistic dispatch, we choose a stepwise reference signal that is updated hourly, and draw the reference for each hourly value from a normal distribution. We choose a mean value of $200\ kW$ for this reference, so that that all devices will be depleted by the end of a single day according to the optimal policy, and set our simulation horizon to 1 day.	We consider two cases with the same mean value: one in which the reference signal values have a high variance and one in which they have a low variance. 
	
\begin{figure}[h]
		\centering
		\includegraphics{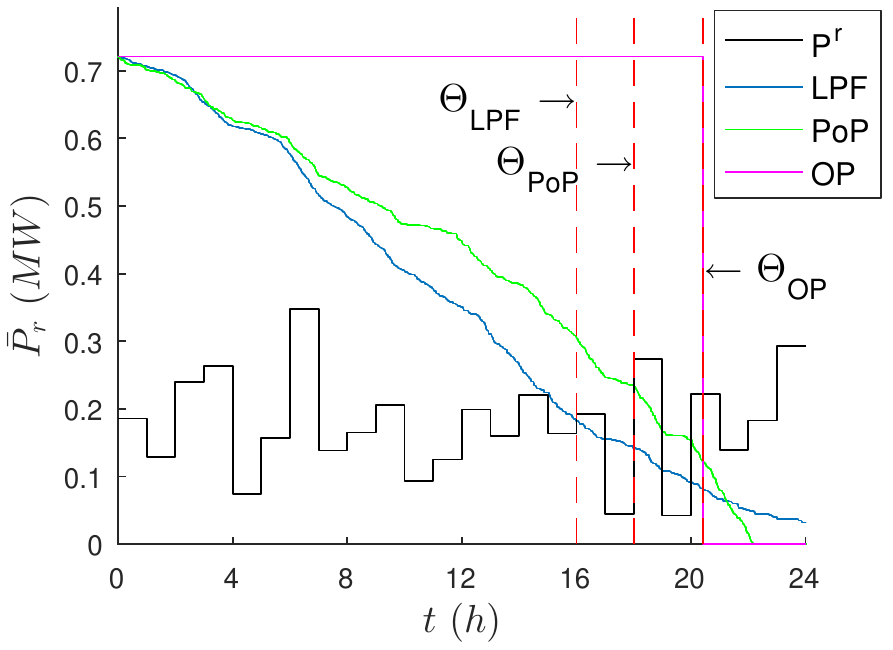}
		\caption{The maximum available power under the implementation of each policy, with the corresponding high-variance reference signal shown for comparison. The time to failure under each policy is highlighted.}
		\label{fig:AvailPlusRefHV}
	\end{figure}

	For the high-variance case, the reference distribution is chosen to be $P^r[k]~\sim~N(200,80)~kW$. In response to a reference signal drawn from this distribution, we plot the evolution of the the maximum available power over time resulting from the implementation of each policy in Figure~\ref{fig:AvailPlusRefHV}. The reference signal is shown for comparison; where the output diverges from this reference is the point of failure of any given policy, as is highlighted. The optimal policy provides the highest feasible reference up to its time to failure, as it postpones emptying devices until absolutely necessary, resulting in the latest time to failure. The practical implications of this result are as follows. If this scenario represents some failure mode requiring less than 16 $h$ for resolution, any of the three policies are capable of maintaining full functionality. If, however, resolution requires between 18 and 21 $h$, it is only the policy we present that is capable of avoiding lost load. Beyond 21 $h$ we are categorically able to say that no policy would be capable of avoiding lost load.
	
	\captionsetup[figure]{font=small, skip=2pt}
	
\begin{figure}[h]
		\centering
		\includegraphics{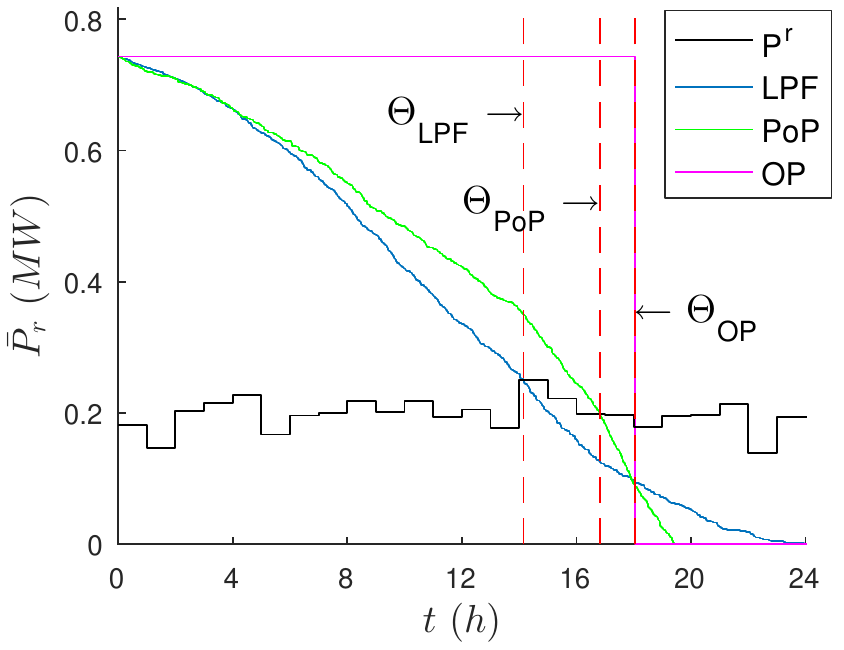}
		\caption{The maximum available power under the implementation of each policy, with the corresponding low-variance reference signal shown for comparison. The time to failure under each policy is highlighted.}
		\label{fig:AvailPlusRefLV}
	\end{figure}

	For the low-variance case, the reference distribution is chosen to be $P^r[k]~\sim~N(200,20)~kW$, and comparative results are shown in Figure~\ref{fig:AvailPlusRefLV}. Once again, the optimal policy results in an improvement over all others, although the increase in time to failure is slightly smaller in this case. This is to be expected, since it is power spikes in particular that cannot be met through the sub-optimal allocation of devices and the resulting depletion of a greater number of them. 
	
	These results show the benefits of the proposed optimal policy in two specific scenarios, but it is worth reiterating that the policy is able to ensure maximum time to failure without prior knowledge of the actual reference signal.
	
	\section{Conclusions and Future Work}
	\label{sec:conc}
	This paper has considered the optimal utilisation of a fleet of energy-constrained distributed resources. We have presented a feedback policy that discharges devices in order of their times-to-go at maximum output power. Our results then prove the optimality of this policy in terms of maximising the time to failure for a given reference signal. We have additionally generalised this optimality into a set-theoretic sense, which is to say that there is no trade-off between different future power reference scenarios; this time-local policy maximises time to failure, available power and future flexibility under any possible future power reference. 
	
	Future work will consider the increased knowledge a system operator or aggregator might acquire by utilising the proposed policy as a proxy for the feasible set. In addition, the authors plan to consider scenarios in which the bi-directional charging capabilities of devices \textit{should} be utilised, including considerations of the additional effects of inefficiencies in these situations.
	
	\section*{Appendix}
	\addcontentsline{toc}{section}{Proofs of Lemmas and Theorem}
	\label{sec:appendix}
	\renewcommand{\theequation}{A.\arabic{equation}}	
	\setcounter{equation}{0}
	\subsection{Proof of \Cref{Lemma III.1}}	
	\proofExistence
	\subsection{Proof of \Cref{Lemma III.5}}
	\proofSupport
	\subsection{Proof of \Cref{Theorem III.7}}
	\proofMaxFlex

	\bibliographystyle{ieeetran}
	\bibliography{biblio,biblio_manual2}
	
\end{document}